\documentstyle[aps]{revtex}

\input{epsf.tex}
\begin{document}
\draft
\twocolumn[\hsize\textwidth\columnwidth\hsize\csname
@twocolumnfalse\endcsname
\renewcommand{\thefootnote}{\alph{footnote}} 
\title{Chaos in Hamiltonians with a Cosmological Time Dependence}
\author{Henry E. Kandrup\footnote{Electronic address: kandrup@astro.ufl.edu}}
\address{ Departent of Astronomy, Department of Physics, and
Institute for Fundamental Theory
\\
University of Florida, Gainesville, Florida 32611}
\date{\today}
\maketitle
\begin{abstract}
This paper summarises a numerical investigation of how the usual 
manifestations of 
chaos and regularity for flows in time-independent Hamiltonians can be 
alterred by a systematic time-dependence of the form arising naturally in an 
expanding Universe. If the time-dependence is not too strong, the observed
behaviour can be understood in an adiabatic approximation. One still infers 
sharp distinctions between regular and chaotic behaviour, even though 
``regular''  does not mean ``periodic'' and ``chaotic'' will not in general 
imply strictly exponential sensitivity towards small changes in initial 
conditions. However, these distinctions are no longer absolute, as it is 
possible for a single orbit to change from regular to chaotic and/or visa 
versa. If the time-dependence becomes too strong, the distinction between 
regular and chaotic can disappear so that no sensitive dependence on initial 
conditions is manifest. 
\end{abstract}
]
\narrowtext
\section{MOTIVATION}
 \label{sec:level1}
In time-independent Hamiltonian systems sharp, qualitative distinctions can be
made between two different types of behaviour, namely regular and chaotic. 
Regular orbits are multiply periodic; chaotic orbits are aperiodic. Chaotic 
orbits exhibit an exponentially sensitive dependence on initial conditions, 
which is manifested by the existence of at least one positive  Lyapunov 
exponent; regular orbits exhibit at most a power law dependence on initial 
conditions\cite{1}. This distinction is, moreover, absolute in the sense that 
it holds for all times: an orbit that starts chaotic will remain chaotic 
forever; a regular orbit remains regular. This distinction is important 
physically because it implies very different sorts of behaviour (although 
topological obstructions like cantori\cite{2} {\it can} make a chaotic orbit 
``nearly regular'' for very long times). 

An obvious question then is to what extent this distinction persists in 
cosmology where, oftentimes, one is confronted with a Hamiltonian manifesting a
systematic time-dependence reflecting the expansion of the Universe. If
the Hamiltonian acquires an explicit secular time-dependence, one would expect
that periodic orbits no longer exist; and one might anticipate further that 
``chaotic'' need not imply a strictly exponential dependence on initial 
conditions. Given, moreover, that the form of the Hamiltonian could change 
significantly over the course of time, one might anticipate the possibility 
that a single orbit could shift in behaviour from ``chaotic'' to ``regular'' 
and/or visa versa. In other words, the distinction between regular and chaotic 
need not be absolute.

Of especial interest is what happens to the gravitational $N$-body problem for
a collection of particles of comparable mass in the context of an expanding 
Universe. It is by now well known\cite{3} that, when formulated 
for a system of compact support which exhibits no systematic expansion, the 
$N$-body problem is chaotic in the sense that small changes in initial 
conditions tend to grow exponentially. Largely independent of the details 
(at least for $N{\;}{\gg}{\;}1$), a small initial perturbation typically
grows exponentially on a time scale $t_{*}{\;}{\sim}{\;}R/v$ comparable to
the natural crossing time for the system. Does this exponential instability 
persist in the context of an expanding Universe, or does the expansion vitiate
the chaos?

To better understand various phenomena in the early Universe, attention has 
also focused on the behaviour of systems modeled as a small number of 
interacting low frequency modes (and, perhaps, an external environment, 
typically visualised as a stochastic bath). The resulting description is 
usually Hamiltonian (albeit possibly perturbed by the external environment), 
but, because of the expansion of the Universe, which induces a systematic 
redshifting of the modes, the Hamiltonian is typically time-dependent. Were 
this time-dependence completely ignored, as might be appropriate in flat space,
the resulting solutions would divide naturally into ``regular'' and 
``chaotic.'' The obvious question is to what extent such appellations continue 
to make sense when one allows properly for an expanding Universe? One might be 
concerned that such systems are intrinsically quantum, and that there is no 
such thing as ``quantum chaos''.\cite{4} However, at least in flat space 
classical distinctions between regular and chaotic behaviour are typically 
manifested in the semi-classical behaviour of true quantum systems, so that 
one might anticipate that any diminution of chaos associated with an expanding 
Universe could have important implications for such phenomena as decoherence 
and the classical-to-quantum transition.\cite{5}

Section II discusses the onset of chaos in time-independent Hamiltonian systems
as resulting from parameteric instability and, by generalising the discussion
to time-dependent Hamiltonians, makes specific predictions as to how chaos 
should be manifested in the context of an expanding Universe. Section III 
confirms and extends these predictions with numerical simulations performed 
for a simple class of models, namely two- and  three-degree-of-freedom 
generalisations of the dihedral potential\cite{6}. Section IV concludes with 
speculations on the implications of these results for real systems in the 
early Universe.

\section{CHAOS AND PARAMETRIC INSTABILITY}
To facilitate sensible predictions as to the manifestations of chaos in 
time-dependent Hamiltonian systems, it is worth
recalling why, in a time-{\it in}dependent Hamiltonian system, chaos 
implies an exponential dependence on initial conditions.

A time-independent Hamiltonian of the form 
\begin{equation}
H({\bf r},{\bf p})={\bf p}^{2}/2 + V({\bf r}) 
\end{equation}
leads immediately to the evolution equation 
\begin{equation}
{d^{2}r^{a}\over dt^{2}}=-{{\partial}V({\bf r})\over {\partial}r^{a}}.
\end{equation}
Whether or not an orbit generated as a solution to this equation is chaotic
depends on how the orbit responds to an infinitesimal perturbation. In
particular, one definition of Lyapunov exponents is that they represent the
average eigenvalues of the linearised stability matrix as evaluated along the
orbit in an asymptotic $t\to\infty$ limit.\cite{7} To understand whether or 
not the orbit is chaotic is thus equivalent to understanding the qualitative 
properties of the linearised evolution equation,
\begin{equation}
{d^{2}{\delta}r^{a}\over dt^{2}}=-{{\partial}^{2}V\over {\partial}r^{b}
{\partial}r^{a}}{\biggr|}_{r_{0}(t)}{\delta}r^{b}. 
\end{equation}

At each point along the orbit, this relation can be viewed as an eigenvalue
equation, with different eigenvectors ${\delta}r^{A}$ satisfying
\begin{equation}
{d^{2}{\delta}r^{A}\over dt^{2}}=-{\Omega}_{A}^{2}(t){\delta}r^{A},
\end{equation}
where ${\Omega}_{A}^{2}$ represents some (in general) complicated function of 
time. Formally, one can expand ${\Omega}_{A}^{2}$ in a complete set of 
modes, both discrete and continuous, so that
\begin{equation}
{d^{2}{\delta}r^{A}\over dt^{2}}=-{\Bigl[} {\cal C}^{A}_{0} + \sum_{\alpha}
{\cal C}^{A}_{\alpha}{\rm cos}\,({\omega}_{\alpha}t+{\varphi}_{\alpha})
{\Bigr]}{\delta}r^{A}, 
\end{equation}
where of course the sum is interpreted as a Stiltjes integral. The important
question is then what sorts of solutions this equation admits.

Were ${\Omega}_{A}^{2}$ constant and negative, this corresponding to an 
imaginary frequency, it is clear that the solution to eq.~(4) would diverge
exponentially, so that the unperturbed orbit must be chaotic. This is in fact
the essence of Hopf's\cite{8} proof that (in modern language) a geodesic flow 
on a compact manifold of constant negative curvature forms a $K$-flow.
However, chaos does not require that ${\Omega}_{A}^{2}$ be negative. Indeed,
as stressed, e.g., by Pettini\cite{9} in a somewhat more formal setting, 
eq.~(4) can admit solutions that grow exponentially even if ${\Omega}_{A}^{2}$ 
is everywhere positive.

As a simple example, suppose that, in eq.~(5), only one time-dependent mode
need be considered, so that (with an appropriate rescaling) this relation 
reduces to the Matthieu equation\cite{10}  
\begin{equation}
{d^{2}{\delta}r^{A}\over dt^{2}}=-({\alpha}+{\beta}{\cos}\,2t){\delta}r^{A}.
\end{equation}
For positive ${\alpha}$ and $|{\beta}|<{\alpha}$, ${\Omega}^{2}$ is always
positive but, nevertheless, for appropriate values of ${\alpha}$ and ${\beta}$
eq.~(6) exhibits a parametric instability which triggers solutions that grow
exponentially in time. 

Eq.~(6) provides a simple way to understand physically why, if one probes a
curve of initial conditions in the phase space of some Hamiltonian system, one 
finds generically that that curve decomposes into disjoint regular and chaotic 
regions. Moving along the phase space curve corresponds to motion through 
the ${\alpha}-{\beta}$ Matthieu plane, but a generic curve in this plane will
typically intersect both stable and unstable regions, i.e., regions where
${\delta}r^{A}$ remains bounded and regions where ${\delta}r^{A}$ grows
exponentially. 

Given this logic, it is easy to see what changes might arise if the unperturbed
evolution equation is altered to allow for the expansion of the Universe. 
Suppose that one is considering a Universe idealised as a spatially flat 
Friedmann cosmology, for which 
\begin{equation}
ds^{2}=-dt^{2}+a^{2}(t){\delta}_{ab}dx^{a}dx^{b}.
\end{equation}
The natural starting point for most field theories is then an action
\begin{equation}
S=\int\,d^{4}x\,a^{3}(t)\,{\biggl[}{\bigl(}{\partial}_{t}{\Phi}{\bigr)}^{2}
-a^{-2}{\delta}^{ab}{\partial}_{a}{\Phi}{\partial}_{b}{\Phi}-V({\Phi})
{\biggr]}
\end{equation}
which leads to mode equations of the form
\begin{equation}
{d^{2}{\phi}_{k}\over dt^{2}}+3{{\dot a}\over a}{d{\phi}_{k}\over dt}=
-{k^{2}\over a^{2}}{\phi}_{k}-{{\partial}V\over {\partial}{\phi}_{k}}.
\end{equation}
Similarly, when formulated in the average co-moving frame, the evolution 
equation for a particle moving in a fixed potential yields a proper peculiar
velocity satisfying\cite{11}
\begin{equation}
{dv^{a}\over dt}+{{\dot a}\over a}v^{a}=-{{\partial}V({\bf r},a(t))\over 
{\partial}r^{a}}.
\end{equation}

In either case, eq.~(2) has been changed in two significant ways, namely
through the introduction of a frame-dragging term ${\propto}\;{{\dot a}/a}$
and the explicit time-dependence now entering into the right hand 
side. The frame-dragging contribution contribution can always be scaled away
by a redefinition of the basic field variable, but the time-dependence on the
right hand side cannot in general be eliminated. It follows that, generically,
eq.~(2) will be replaced by
\begin{equation}
{d^{2}r^{a}\over dt^{2}}=-{{\partial}V({\bf r},a(t))\over {\partial}r^{a}}.
\end{equation}
The simplest case arises when the time-dependence enters only as an overall 
multiplicative factor, so that
\begin{equation}
{d^{2}r^{a}\over dt^{2}}=-R[a(t)]{{\partial}V({\bf r})\over {\partial}r^{a}}.
\end{equation}

To the extent that perturbations of a solution to eq.~(2) can be represented
qualitatively by the Matthieu eq.~(6), it is not unreasonable to suppose that
perturbations of its time-dependent generalisation (12) can be represented 
by a generalised Matthieu equation of the form
\begin{equation}
{d^{2}{\delta}r^{A}\over dt^{2}}=
-R(t)({\alpha}+{\beta}{\cos}\,2t){\delta}r^{A}.
\end{equation}
or, perhaps, some generalisation thereof in which $2t$ is replaced by some
$T(t)$.

The qualitative character of the solutions to this equation are easy to predict
theoretically and corroborate numerically, at least when the time-dependence
of $R$ is not too strong and this time variability can be treated in an 
adiabatic approximation: There is still a sharp distinction between stable 
and unstable behaviour, but ``unstable'' does not in general correspond to 
strictly exponential growth. For the true Matthieu equation (6), unstable
solutions {\it do} grow exponentially, with 
${\delta}r^{A}{\;}{\propto}{\;}{\exp}({\omega}t)$, but allowing for a 
nontrivial $R(t)$ leads instead to solutions of the form
\begin{equation}
{\delta r}^{A}{\;}{\sim}{\;}{\exp}\,[\int \,R^{1/2}(t){\omega}dt]
\end{equation}
If, e.g., $R(t){\;}{\propto}{\;}t^{p}$, chaos should correspond to the 
existence of perturbations that grow as 
\begin{equation}
{\delta r}^{A}{\;}{\sim}{\;}{\exp}[t^{1+p/2}].
\end{equation}
In other words, the evolution of ${\delta}r^{A}$ will be sub- or 
super-exponential. 

Two other points should also be clear. First and foremost, the distinction
between regular and chaotic should no longer be absolute. Equation (13) can
be interpreted as an ordinary Matthieu equation with time-dependent 
``constants'' ${\hat\alpha}=R^{1/2}{\alpha}$ and ${\hat\beta}=R^{1/2}{\beta}$. 
However, the fact that these ``constants'' change in time means that the 
equation satisfied by ${\delta}r^{A}$ is drifting through the 
${\alpha}-{\beta}$ plane, so that the solutions can drift into and/or out of
resonance. In other words, the behaviour of a small perturbation can in 
principle change from stable to unstable and/or visa versa, which corresponds
to the possibility of transitions between regular and chaotic behaviour.

The other point is that, if the time-dependence is too strong, the adiabatic
approximation could fail and the expected behaviour could be quite different. 
In particular, for $R{\;}{\propto}{\;}t^{-p}$, with $p\to -2$, eq.~(15) 
implies that the subexponential evolution expected for $p$ somewhat smaller
than zero will degenerate into a simple power law evolution where, seemingly,
all hints of chaos are lost. An important question would seem to be how 
negative $p$ must become before the distinctions between chaos and regularity 
are completely obliterated.

Corroboration of the behaviour predicted in eq.~(15) and a partial answer to 
this last question are provided by the numerical computations described in the 
following section.

\section{NUMERICAL SIMULATIONS}

The experiments described here were performed for time-dependent extensions
of the potential 
$$V_{0}(x,y,z)=-(x^{2}+y^{2}+z^{2})+{1\over 4}(x^{2}+y^{2}+z^{2})^{2}$$
\begin{equation}
-{1\over 4}(y^{2}z^{2}+bz^{2}x^{2}+cx^{2}y^{2}),
\end{equation}
with variable parameters $b$ and $c$ of order unity, which constitutes a
three-dimensional analogue of the so-called dihedral potential\cite{6}. For
$b=c=1$, the case treated in greatest detail, this corresponds to a slightly 
cubed Mexican hat potential. At high energies, the potential is essentially
quartic, although the quadratic couplings ensure that there are significant 
measures of both regular and chaotic orbits. At relatively low energies, 
somewhat less than zero, orbits are confined to a three-dimensional trough 
where, qualitatively, they behave like orbits in the ``stadium 
problem,''\cite{12} scattering off the walls in a fashion that renders them 
largely chaotic.

Some of the experiments involved allowing for a time-dependence of the form
\begin{equation}
V(x,y,z,t)=R(t)V_{0}(x,y,z)
\end{equation}
with $R(t){\;}{\propto}{\;}t^{p}$. Others involved the alternative
time-dependence 
\begin{equation}
V(x,y,z,t)=V_{0}[R(t)t,R(t)y,R(t)z],
\end{equation}
again with $R(t){\;}{\propto}{\;}t^{p}$.

Individual sets of experiments involved freezing the energy at some fixed
value, usually $E[V_{0}]=10.0$, and selecting an ensemble of some $1000$ to
$5000$ initial conditions. Three-degree-of-freedom initial conditions were 
generated by setting $x=0$, $z=$ const, uniformly sampling the energeticaly 
allowed regions of the $y-z-p_{y}-p_{z}$ hypercube, and then solving for 
$p_{x}=p_{x}(y,z,p_{y},p_{z},E)$. Two-degree-of-freedom initial conditions were
generated by setting $x=z=p_{z}=0$, uniformly sampling the energetically 
allowed regions of of $y-p_{y}$ plane, and solving for $p_{x}$
$=p_{x}(y,p_{y},E)$. The computations were started at $t=1$, $t=10$, or 
$t=100$ and ran for a total time $T=256$ or $T=512$, this corresponding in the
absence of any time-dependence to ${\sim}{\;}100-200$ characteristic crossing
times. The orbital data were recorded at intervals ${\Delta}t=0.25$. An
estimate of the largest short time Lyapunov exponent\cite{13} was computed in
the usual way\cite{6} by simultaneously evolving a slightly perturbed orbit
(initial perturbation ${\delta}x=1.0\times 10^{-8}$) which was periodically
renormalised at intervals ${\Delta}t=1.0$ to keep the perturbation small.
This led for each orbit to a numerical approximation to the quantity
\begin{equation}
{\chi}(t)=\lim_{{\delta}Z(0)\to 0}\,{1\over t}\;{\ln}\,{\Biggl[}
{||{\delta}Z(t)||\over ||{\delta}Z(0)||} {\Biggr]}
\end{equation}
for a phase space perturbation ${\delta}Z=(|{\delta}{\bf r}|^{2}+
|{\delta}{\bf p}|^{2}).^{1/2}$

Since interest focuses on the probability that ``chaos'' does not correspond
to a purely exponential growth, the data were reinterpreted by partitioning
the Lyapunov data into bits of length ${\Delta}t=1.0$, each of which probed
the growth of the perturbation during the interval ${\Delta}t_{i}$:
\begin{equation}
{\chi}({\Delta}t_{i})={{\chi}(t_{i}+{\Delta}t)(t_{i}+{\Delta}t)-
{\chi}(t_{i})t_{i}\over {\Delta}t}.
\end{equation}
These bits were recombined to generate partial sums
\begin{equation}
{\xi}(t_{i})={1\over {\Delta}t}\;
\sum_{j=1}^{i-1}\,{\chi}({\Delta}t_{j})
={1\over {\Delta}t}\,  {\ln}\,{\Biggl[}
{||{\delta}Z(t_{i}+t_{0})||\over ||{\delta}Z(t_{0})|||} {\Biggr]},
\end{equation}
which capture the net growth of the logarithm of the perturbation. These 
partial sums were then fit to a polynomial growth law
\begin{equation}
{\xi}=A+Bt^{q}
\end{equation}
A purely exponential growth would yield $q=1$; sub- and super-exponential
growth correspond, respectively, to $q<1$ and $q>1$. If the perturbation only
grows as a power law, eq.~(22) should not yield a reasonable fit.

The principal conclusion of the computations is that, overall, the adiabatic
approximation works very well so that, for a broad range of values of $p$ in
the potential (17), eq.~(15) appears satisfied. This is illustrated in Fig.
1 (a), which exhibits the best fit exponent $q$ as a function of $p$. The 
data in this plot combine experiments for both two- and three-degree-of-freedom
chaotic orbits which allowed for several different values of $b$ and $c$. 
Changing the values of $b$ and/or $c$ can significantly change the rate at 
which a small initial perturbation grows, thus altering the characteristic 
value of $B$ in eq.~(22), but the exponent $q$ seems independent of these 
details. 

As a concrete example, consider the special case of two-degree-of-freedom
orbits. For $p=0$, i.e., no time-dependence, initial conditions
with $E[V_{0}]=10.0$ evolved in (17) exhibit a clean split into regular and 
chaotic, with some 72\% of the orbits chaotic and the remaining regular. 
Moreover, for this initial energy cantori are relatively are unimportant, so 
that there are few ``sticky'' chaotic orbits trapped near regular islands. 
$N[{\xi}]$, the distribution of the final values of ${\xi}$, is strongly  
bimodal, and it is almost always easy to distinguish visually between 
regularity and chaos. For the chaotic orbits one finds (as must be the case) 
excellent agreement with eq.~(15).

As $p$ increases to assume small positive values, the final ${\xi}$'s continue 
to yield a bimodal distibution, $N[{\xi}]$, which indicates that, in terms of 
their sensitive dependence on initial conditions, the orbits still divide into 
two distinct classes. However, there is a systematic decrease in 
the abundance of regular orbits, so much so that, for $p>0.3$, there are few 
if any regular orbits. Moreover, a detailed examination of individual orbits 
show that they can exhibit abrupt changes in behaviour between chaotic 
intervals where ${\xi}$ grows as $t^{1+p/2}$ and regular intervals where 
${\xi}$ exhibits little if any growth. For $0.3<p<0.6$ (almost) every orbit in 
the evolved ensembles of initial conditions seems a member of a single
chaotic population with $q=1+p/2$. 

However, for larger values of $p$ orbits which began as chaotic subsequently
exhibit abrupt transitions to regularity and remain regular for the duration
of the integration. This behaviour reflects that fact that, at sufficiently
late times, orbits with $p>0$ become trapped in one of the four global
minima of the potential with $V_{0}(x_{0},y_{0})=-4/3$, located at $x_{0}=
{\pm}{\sqrt{ 4/3}}$, $y_{0}={\pm}{\sqrt{ 4/3}}$, and oscillates in what is 
essentially an integrable quadratic potential. Indeed, when the orbit becomes 
trapped close enough to one of these minima, so that $V$ is strictly negative, 
the orbit quickly loses energy until it comes to sit almost exactly at rest. 
This implies that, eventually, ${\xi}$ decreases with time, this reflecting 
the fact that all orbits in the given basin of attraction are driven towards 
the same final point.

As $p$ decreases from zero to somewhat negative values, two more or less
distinct populations again appear to persist, at least initially, although now 
the relative abundance of chaotic orbits decreases rapidly. For relatively
small values of $p$, say $p<-0.5$ or so, ${\xi}$ grows so slowly that the
chaotic and regular contributions to $N[{\xi}]$ exhibit some considerable
overlap, and it is no longer easy in every case to distinguish regular from
chaotic. For $p<-1.0$ the relative measure of chaotic orbits appears to be
extremely small, and for $p<-1.8$ it is unclear whether any ``chaotic'' orbits 
exist at all. To the extent that ``chaotic orbits'' do exist, they are nearly
indistinguishable from ``regular'' orbits. 

These general trends are manifested in Figs.~2 (a) - (f), which exhibits the 
final $N[{\xi}]$ at $t=266$ for integrations with, respectively, $p=-1.8$, 
$p=-0.5$, $p=0.0$, $p=0.2$, $p=0.5$, and $p=2.5$. Further insights into the 
behaviour of small perturbations can be inferred from Figs.~3 (a) - (f), which 
plot ${\xi}(t)$ for $150$ randomly chosen orbits from each of these 
integrations. 

This behaviour can be contrasted with what obtains for two-degree-of-freedom
orbits for the same potential $V_{0}$ if one now allows for the time-dependence
given by eq.~(18). Here once again one finds that changing $p$ leads to 
sub- or super-exponential sensitivity, and, as is evident from Fig. 1~(b), 
that increasing $p$ tends to yield larger values of $q$. Moreover, there is 
often tangible evidence for two distinct types of orbits, seemingly chaotic 
and regular. However, the details change appreciably.

In this case, one finds that increasing $p$ from zero to slightly
positive values tends to increase the overall abundance of regular orbits, and
that even those orbits which seems chaotic overall often exhibit regular
intervals during which ${\xi}$ exhibits little, if any, growth. For values
as alrge as $p=0.4$, it seems that most -- albeit not all -- of the orbits
are regular nearly all the time. However, for somewhat larger values of $p$
the relative importance of chaos again begins to increase, although one 
continues to observe regular intervals. For sufficiently high values of $p$, 
one finds that, as for the potential (17), all the orbits eventually get 
trapped near one of the four minima of the potential, at which point the 
orbits become completely regular and ${\xi}$ becomes negative.

Alternatively, if one passes from zero to negative values of $p$, one finds
that the relative abundance of regular orbits rapidly decreases. Indeed, for
$p<-0.1$ or so there appear to be virtually no regular orbits. However, the
chaos is vitiated in the sense that the growth of a small perturbation is
decidedly subexponential. Indeed, as $p$ decreases, one quickly reaches a 
point where the dependence on initial conditions is so weak that, even though
it seems reasonable to term the orbits chaotic, that chaos must be 
viewed as extremely weak.

This behaviour is exhibited in Figs.~4 (a) - (f), which plot $N[{\xi}]$ at 
$t=266$ for integrations with, respectively, $p=-0.6$, $p=-0.3$, $p=-0.05$, 
$p=0.5$. $p=0.9$ and $p=1.25$. Figs.~3 (a) - (f) plot ${\xi}(t)$ for $150$ 
randomly chosen orbits from each of these integrations. 

In terms of their sensitive dependence on initial conditions orbits in these 
time-independent Hamiltonian systems tend to divide relatively cleanly into 
two distinct classes; and, for the case of a time-dependence of the form given 
by eq.~(17), those that exhibit such a sensitive dependence are well fit 
overall by the scaling predicted by eq.~(15). However, to justify designating
these orbit classes ``regular'' and ``chaotic,'' it is important to 
verify that these distinctions also coincide with the general shapes of the 
orbits, as manifested visually or through the computation of a Fourier 
transform. 

This is in fact easily done. Orbits that are chaotic in terms of their
sensitive dependence on initial conditions tendy systematically to look ``more
irregular'' and to have ``messier'' Fourier spectra than do those which do
not manifest a sensitive dependence on initial conditions. For time-independent
Hamiltonian systems, one can distinguish between regular and chaotic by 
determining the extent to which most of the power is concentrated in a few
special frequencies: in a $t\to\infty$ limit, the power spectrum for a regular
orbit will only have support at a countable set of frequencies whereas a 
chaotic orbit will typically have power for a continuous range of frequencies.
However, this is {\it not} the case for a time-dependent Hamiltonian system! 
As the energy changes in time, the characteristic frequencies of an orbit 
must change so that a long time integration necessarily yields broad band 
power. However, what {\it is} true for a regular orbit is that this power
tends to vary comparatively smoothly with frequency. 

This can be understood easily in the context of an adiabatic approximation.
At any given instant, it makes sense to speak of the two principal frequencies 
that define (say) a two-degree-of-freedom loop orbit; but, as the energy
changes, the values of these two frequencies will change continuously with
time. Indeed, the phase space can eventually evolve to the point that an orbit
that starts as a loop becomes unstable and is transformed into a chaotic orbit
or, perhaps, a different type of regular orbit.

Representative examples of regular and chaotic orbits in time-dependent
Hamiltonians are given in Figs. 6 and 7 which, respectively, exhibit data
generated for the potential (17) with $p=-0.4$ and the potential (18) with
$p=0.3$.\cite{14} In each Figure, the left hand panels correspond to 
two-degree-of-freedom chaotic orbits; the right hand sides correspond to
two-degree-of-freedom regular orbits. The top panels display the orbits in
the $x-y$ plane. The middle panels exhibit the power spectra, $|x({\omega})|$
and $|y({\omega})|$. The bottom panels show the total ${\xi}(t)$.

\section{PHYSICAL IMPLICATIONS}

At least for the time-dependent potentials described in this paper, it seems
possible to make sharp distinctions between regularity and and chaos,
even if these distinctions are not absolute: as its energy changes, an orbit
can evolve from regular to chaotic and/or visa versa. However, the
time-dependence alters the exponential dependence on initial conditions
manifested in the absence of any time-dependence, yielding instead a sub- or
superexponential sensitivity. For the models described in Section III, $p<0$
corresponds physically to an expanding Universe; and, as such, the computations
corroborate the physical expectation that the expansion should suppress
chaos. Whether or not in the real Universe this expansion is strong enough
to kill the chaos completely depends on the details of the assumed behaviour
of both the scale factor and the dynamical model.

But why should one care? Why would the presence or absence of chaos 
matter? Perhaps the most important implication of chaos for the bulk evolution
of self-gravitating systems is its potential role in {\it violent relaxation},
the coarse-grained evolution of a non-dissipative system towards a well-mixed 
state.
As formulated originally by Lynden-Bell,\cite{15} violent relaxation is a 
phase mixing process whereby a generic blob of collisionless matter, be it
localised in space or characterised by any other phase space distribution, 
tends to disperse until it approaches some equilibrium or steady state. 
The crucial question is one of efficiency. How quickly, and how
efficiently, will the matter disperse? At a given level of resolution, how long
must one wait before the matter constitutes a reasonable sampling of a
near-equilibrium?

The important point, then, is that the answer to these questions depends 
crucially on whether the flow be regular or chaotic.\cite{16} If, in the 
absence of an expanding Universe, the matter executes regular, nonchaotic 
trajectories, the approach towards equilibrium will be comparatively 
inefficient. As probed by a coarse-grained distribution function and/or lower 
order phase space moments, there {\it is} a coarse-grained evolution towards 
equilibrium, but it proceeds only as a power law in time. If, however, the
matter executes chaotic orbits, this approach will be exponential in time,
proceeding at a rate that correlates with the values of the positive Lyapunov 
exponents.

The obvious inference is that to the extent that the expansion 
of the Universe weakens chaos, it should also weaken this 
chaotic mixing. This suggests that phase mixing will be a comparatively 
inefficient mechanism at early times, when  the dynamics is dominated by the
overall expansion, and that it can only begin to play an important role at
later times, once an overdense region has ``pinched off'' from the overall
expansion and begun to evolve more or less independently. 

\acknowledgments
The author would like to acknowledge arguments with himself, most of which he
lost.

\pagestyle{empty}
\begin{figure}[t]
\centering
\centerline{
        \epsfxsize=8cm
        \epsffile{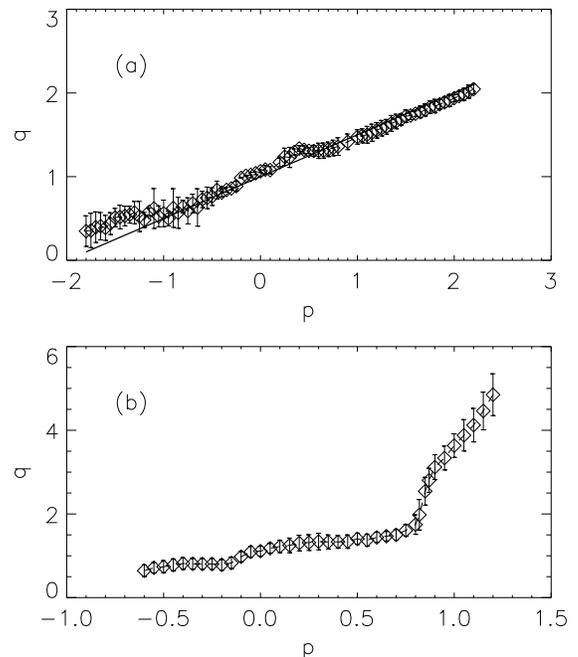}
           }
        \begin{minipage}{10cm}
        \end{minipage}
        \vskip -0.3in\hskip -0.0in
\caption{(a) Best fit exponent $q$ for a growth rate 
${\propto}{\;}\exp(at^{q})$ for two-degree-of-freedom chaotic orbits evolving 
in the potential $V=t^{p}V_{0}({\bf r})$ with $a=b=1$. The line yields the 
prediction of an adiabatic approximation. (b) The same, except for the 
potential $V=V_{0}(t^{p}{\bf r})$.}
\vspace{-0.0cm}
\end{figure}
\vfill\eject
\pagestyle{empty}
\begin{figure}[t]
\centering
\centerline{
        \epsfxsize=8cm
        \epsffile{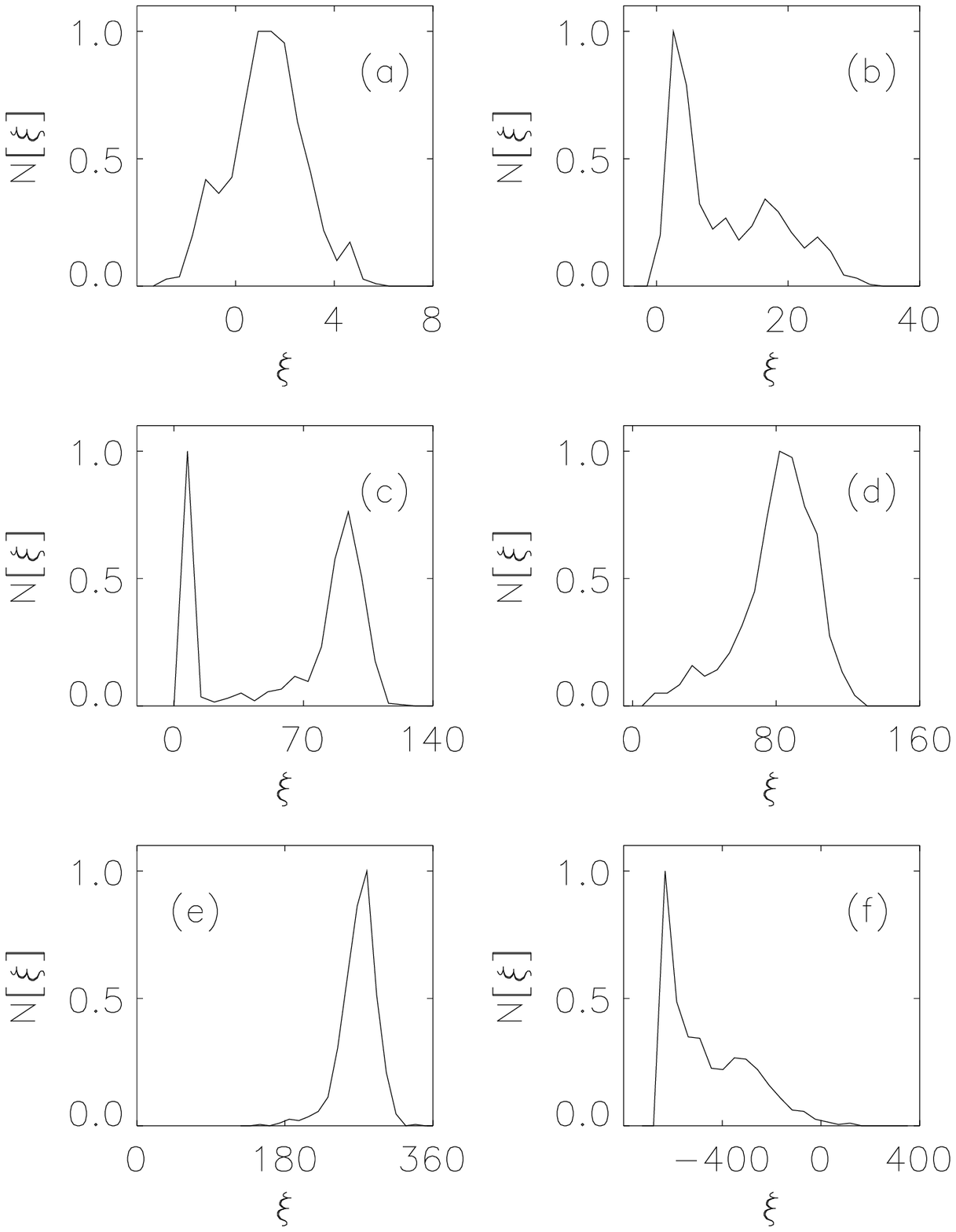}
           }
        \begin{minipage}{10cm}
        \end{minipage}
        \vskip -0.0in\hskip -0.0in
\caption{(a) Normalised distribution $N[{\xi}(t=522)]$ for 
two-degree-of-freedom orbits with $a=b=1$ evolved for the interval $10<t<266$
in the potential $V=V_{0}(t^{p}{\bf r})$ with $p=-1.8$. (b) The same for 
$p=-0.5$. (c) $p=0.0$. (d) $p=0.2$. (e) $p=0.5$ (f) $p=2.5$.}
\vspace{0.0cm}
\end{figure}

\pagestyle{empty}
\begin{figure}[t]
\centering
\centerline{
        \epsfxsize=8cm
        \epsffile{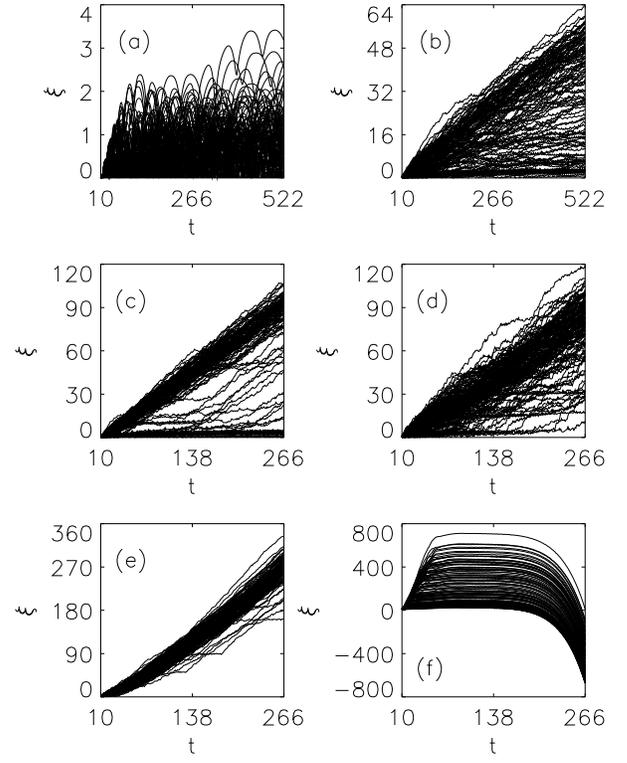}
           }
        \begin{minipage}{10cm}
        \end{minipage}
        \vskip -0.0in\hskip -0.0in
\caption{(a) ${\xi}(t)$ for $150$ representative two-degree-of-free- dom orbits 
evolved in the potential $V=t^{p}V_{0}({\bf r})$ with $p=-1.8$ (b) The same
for $p=-0.5$. (c) $p=0.0$. (d) $p=0.2$. (e) $p=0.5$ (f) $p=2.5$.}
\vspace{0.0cm}
\end{figure}

\pagestyle{empty}
\begin{figure}[t]
\centering
\centerline{
        \epsfxsize=8cm
        \epsffile{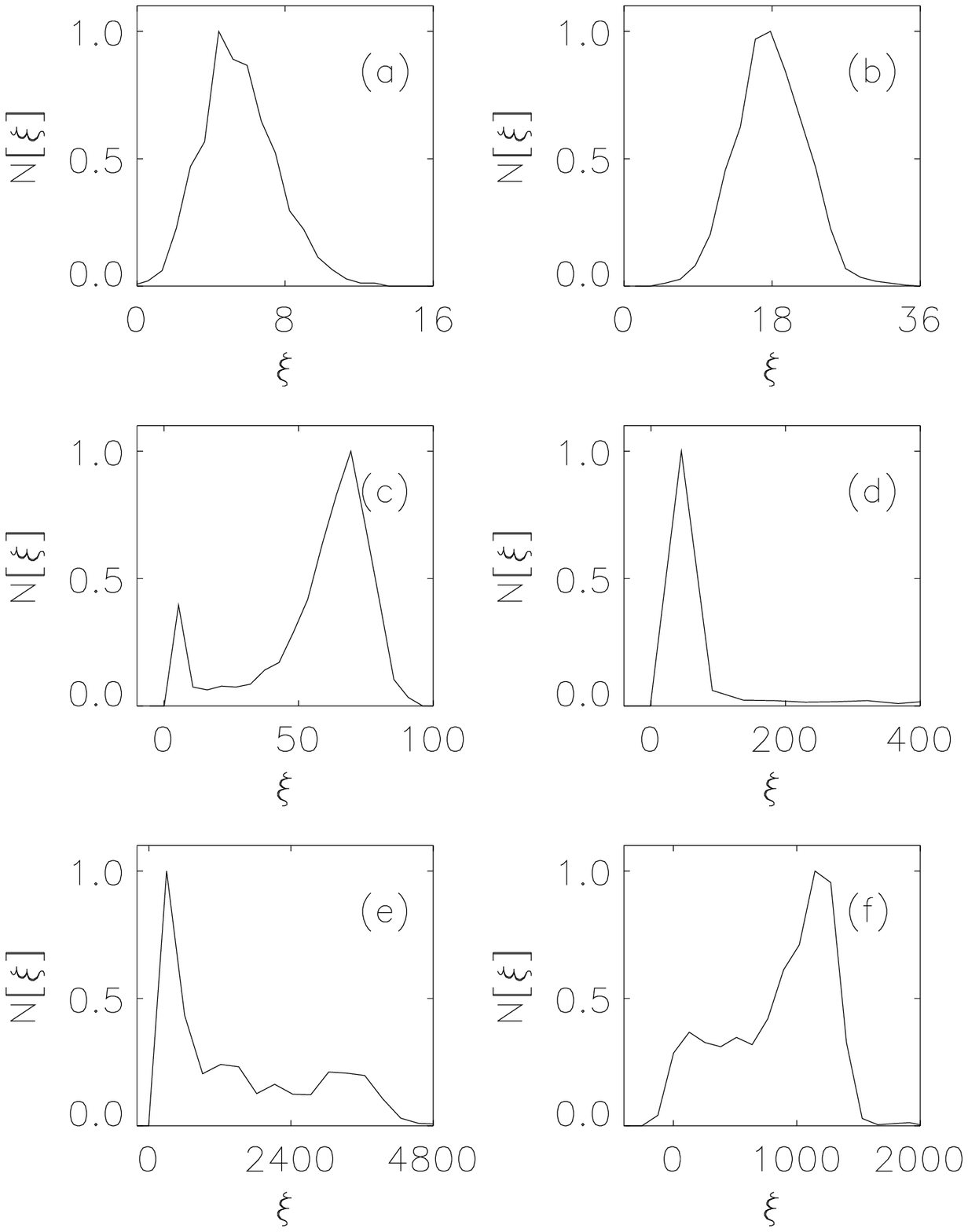}
           }
        \begin{minipage}{10cm}
        \end{minipage}
        \vskip -0.0in\hskip -0.0in
\caption{(a) Normalised distribution $N[{\xi}(t=522)]$ for 
two-degree-of-freedom orbits with $a=b=1$ evolved for the interval $10<t<266$
in the potential $V=V_{0}(t^{p}{\bf r})$ with $p=-0.6$. (b) The same for 
$p=-0.3$. (c) $p=-0.05$. (d) $p=0.5$. (e) $p=0.9$ (f) $p=1.25$.}
\vspace{0.0cm}
\end{figure}

\pagestyle{empty}
\begin{figure}[t]
\centering
\centerline{
        \epsfxsize=8cm
        \epsffile{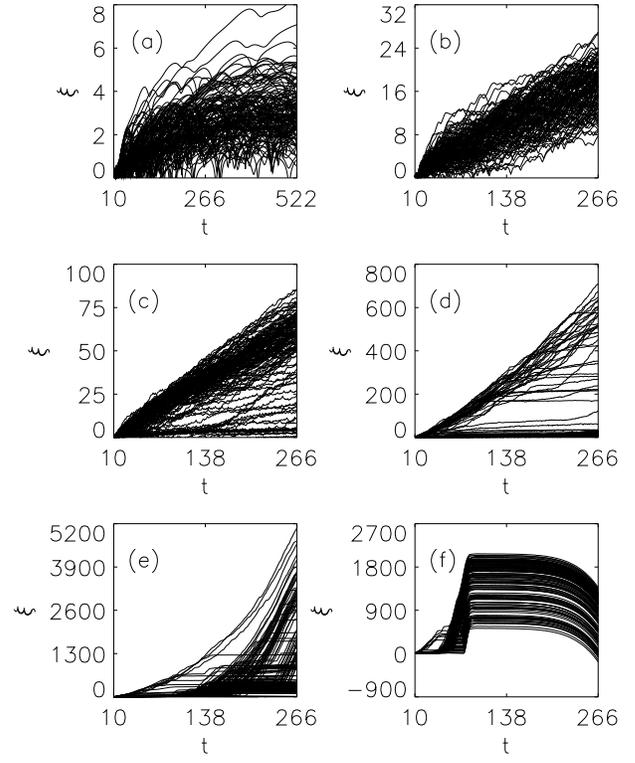}
           }
        \begin{minipage}{10cm}
        \end{minipage}
        \vskip -0.0in\hskip -0.0in
\caption{(a) ${\xi}(t)$ for $150$ representative two-degree-of-free- dom orbits 
evolved in the potential $V=V_{0}(t^{p}{\bf r})$ with $p=-0.6$ (b) The same
for $p=-0.3$. (c) $p=-0.05$. (d) $p=0.3$. (e) $p=0.9$ (f) $p=1.25$.}
\vspace{0.0cm}
\end{figure}

\pagestyle{empty}
\begin{figure}[t]
\centering
\centerline{
        \epsfxsize=8cm
        \epsffile{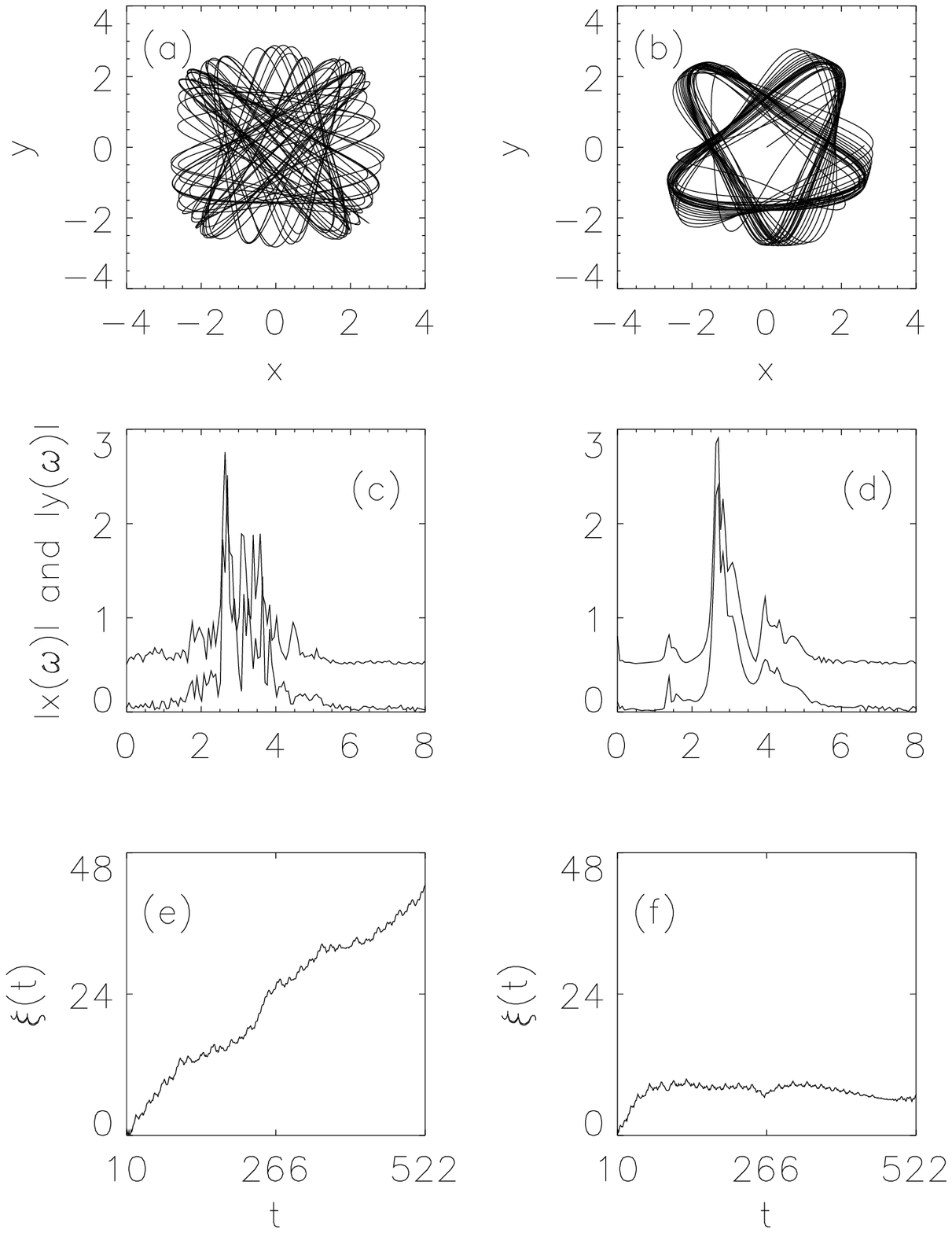}
           }
        \begin{minipage}{10cm}
        \end{minipage}
        \vskip -0.0in\hskip -0.0in
\caption{(a) A chaotic two-degree-of-freedom orbit evolved in the potential 
(17) with $p=-0.4$. (b) A regular orbit evolved with the same value of $p$.
(c) The Fourier transformed quantities $10^{-3}|x({\omega})|$ and 
$10^{-3}|y({\omega})|$ (the latter translated upwards by $0.5$) generated from 
the orbit in (a). (d) The same for the orbit in (b). (e) ${\xi}(t)$ for the
orbit in (a). (f) The same for the orbit in (b).}
\vspace{0.0cm}
\end{figure}

\pagestyle{empty}
\begin{figure}[t]
\centering
\centerline{
        \epsfxsize=8cm
        \epsffile{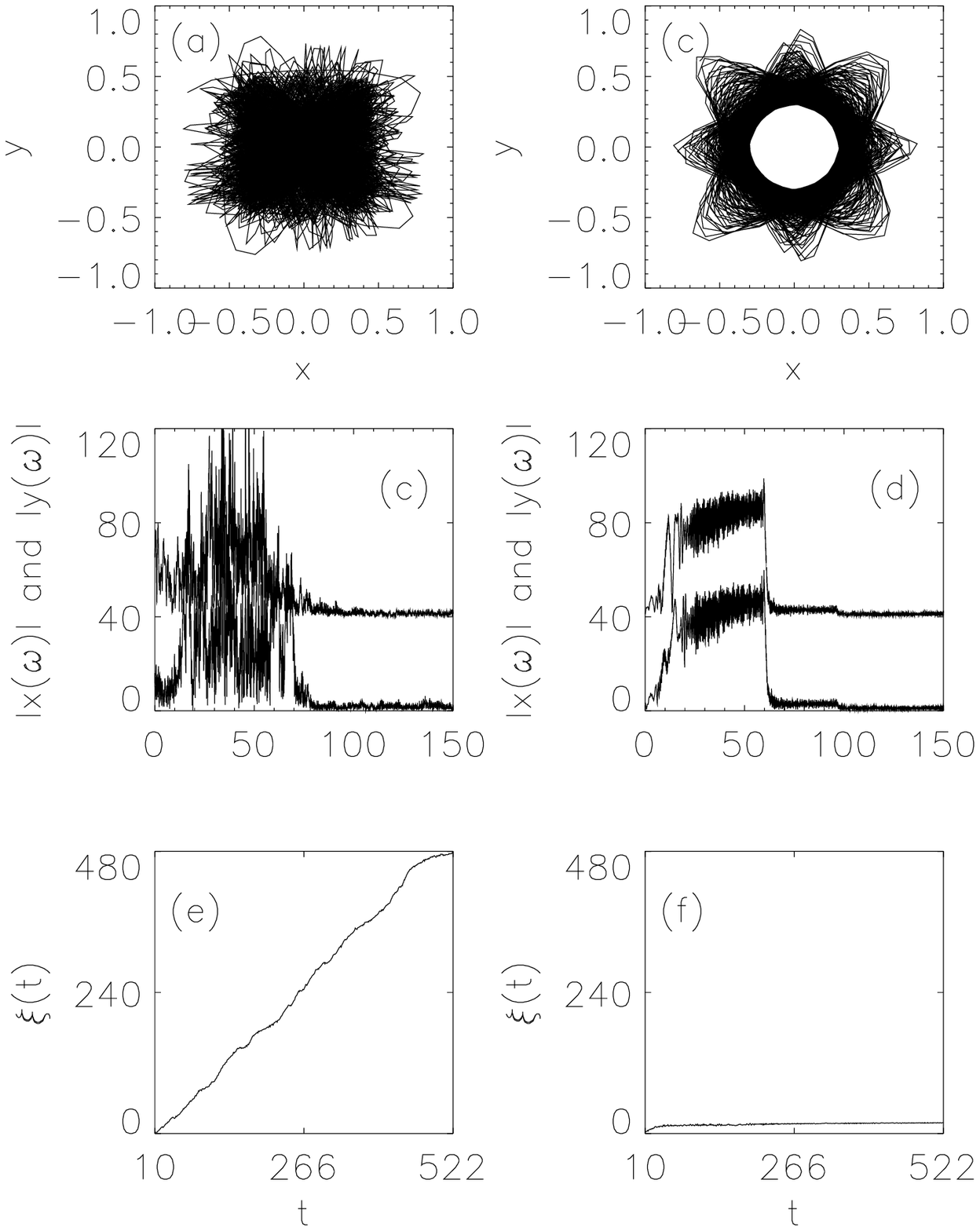}
           }
        \begin{minipage}{10cm}
        \end{minipage}
        \vskip -0.0in\hskip -0.0in
\caption{(a) A chaotic two-degree-of-freedom orbit evolved in the potential 
(18) with $p=0.3$. (b) A regular orbit evolved with the same value of $p$.
(c) The Fourier transformed quantities $|x({\omega})|$ and 
$|y({\omega})|$ (the latter translated upwards by $40$) generated from 
the orbit in (a). (d) The same for the orbit in (b). (e) ${\xi}(t)$ for the
orbit in (a). (f) The same for the orbit in (b).}
\vspace{0.0cm}
\end{figure}
\vfill\eject

\begin{references}
\bibitem{14} The regular examples in Figs. 6 and 7 both correspond to loop
orbits, which are characterised by a net sense of rotation. However, with or 
without a time-dependence these potentials also admit large numbers of so 
called box orbits which exhibit no net sense of rotation and which, 
topologically, resmeble Lissajous figures.
\end{references}
\end{document}